\documentclass[prb,twocolumn,showpacs,preprintnumbers,amsmath,amssymb,superscriptaddress]{revtex4}
\usepackage{epsf}
\usepackage{graphicx}
\usepackage{bm}
\usepackage{amsmath}
\usepackage{color}
\usepackage{notes2bib}
\usepackage{epstopdf}

\bibnotesetup{
note-name = , use-sort-key = false }
\begin{document}

\title{Spin-dependent tunneling in semiconductor heterostructures with a magnetic layer}

\author{I.~V.~Rozhansky}
\email{rozhansky@gmail.com} \affiliation{Ioffe Institute, 194021
St.Petersburg, Russia} \affiliation{Lappeenranta University of
Technology, FI-53851 Lappeenranta, Finland}
\affiliation{Peter the Great Saint-Petersburg Polytechnic University, 195251 St. Petersburg, Russia}
\author{K.~S.~Denisov}
\affiliation{Ioffe Institute, 194021 St.Petersburg, Russia}
\author{N.~S.~Averkiev}
\affiliation{Ioffe Institute, 194021 St.Petersburg, Russia}
\author{E.~L\"ahderanta}
\affiliation{Lappeenranta University of Technology, FI-53851 Lappeenranta, Finland}

\begin{abstract}
We present a theory that describes the appearance of circular polarization of the photoluminescence (PL) in ferromagnet-semiconductor hybrid heterostructures due to spin-dependent tunneling of photoexcited carriers from a quantum well into a magnetic layer.
The theory succeeds in explaining the experimental data on time-resolved PL for heterostructures
consisting of InGaAs-based quantum well (QW) and a spatially separated Mn $\delta$-layer.
We show that the circular polarization of the PL originates from dynamic spin polarization
of electrons due to spin-dependent leakage from the QW onto Mn donor states split by the exchange field
of the ferromagnetic Mn delta-layer.
\end{abstract}

\pacs{78.20.Ls, 
78.67.-m, 
75.50.Pp, 
75.76.+j 
 }

\date{\today}

\maketitle

\section{Introduction}

The field of semiconductor spintronics can be now claimed as well-established. However,
the 'classical' spintronic devices such as spin transistor \cite{Datta} or spin valves \cite{spinValve}
still do not meet the theoretical expectations to advance the modern applied electronics.
The key issue yet to be resolved along the way is the fabrication of a good semiconductor with  ferromagnetic properties. A substantial breakthrough was the discovery of (Ga,Mn)As dilute magnetic semiconductor (DMS) \cite{Ohno} with relatively high Curie temperature of around $T_c\approx100$ K. The highest Curie temperature achieved for bulk dilute (Ga,Mn)As samples does not exceed $200$ K so far \cite{Dietl2014}.
While the Mn solubility limit basically prevents further increase of $T_c$ in bulk samples, the hybrid (Ga,Mn)As
heterostructures with Mn layer coupled to a remote 2D holes channel have gained a considerable interest
\cite{Rupprecht,Nishitani,Aronzon2010}.
The GaAs-based heterostructure with a Mn $\delta$-layer located in a vicinity of In$_x$Ga$_{1-x}$As quantum well (QW) exibits a ferromagnetic behavior similar to that of the bulk Mn-doped GaAs DMS.
It was demonstrated that the 2D holes populating the QW substantially contribute to the
ferromagnetism of the Mn layer due to resonant indirect exchange interaction \cite{OURPss2014,ourAPL2015}.

  The ferromagnetic ordering of the Mn $\delta$-layer also gives rise to the circular polarization of the
photoluminescence (PL) from the QW. However, the particular microscopic mechanism
leading to this phenomena still has not been fully understood.
Ferromagnetic (Ga,Mn)As DMS are p-type semiconductors so the logical assumption might be that
in the thermal equilibrium the QW is populated with 2D holes which are spin-polarized due to coupling with Mn ions
so that the light emitted from the QW would be circular polarized. The theory of this mechanism has been developed in \cite{OURPRB2013,OURLowTempPhys}
and it is probably relevant to the experimental data reported in \cite{Kulakovskii,Zaitsev}.
However the recent time-resolved experiments on similar samples with more shallow In$_x$Ga$_{1-x}$As QW
have demonstrated that under moderate photoexcitation the spin polarization in GaMnAs-based hybrid is a non-equilibrium, dymanic effect \cite{KorenevSapega}.
 We argue that in these experiments the circular polarization of the PL stems from the dynamic spin polarization of the photoexcited electrons.
This finding does not completely exclude the importance of the holes spin polarization,
but only states the prime role of the dynamic electrons in the hybrid structures with a shallow QW (and far less sheet density of the 2D holes that in the samples with deeper QW).
 In the present paper we provide a theory for the effect which perfectly describes the experimental data.
In our model the spin polarization of the electrons remaining in the QW occurs due to an effective spin-dependent tunneling into the magnetic layer
followed by a non-radiative recombination. 
 The theoretical description developed appears to be rather general
 and can be applied to various semiconductor heterostructures with a similar design.

\section{The model}
\begin{figure}
 \centering\includegraphics[width=0.4\textwidth]{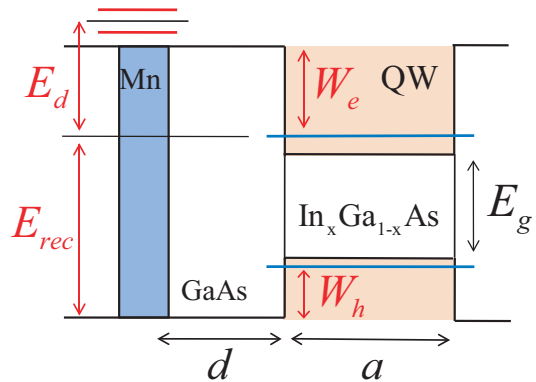}
 \caption{Schematic band structure of the considered ferromagnetic-semiconductor system.}
 \label{fig1}
\end{figure}

 The band diagram of the system under study is shown schematically in Fig.\ref{fig1}. It consists of In${_x}$Ga$_{1-x}$As QW sandwiched between GaAs barriers
and a thin layer doped with Mn located at a few nanometers from the QW. The Mn layer comprises a dilute magnetic
semiconductor with a pronounced ferromagnetic behavior.
Our theory focuses on the electrons tunneling from the QW into the ferromagnetic Mn layer.
The theoretical description developed further is rather general
 and can be applied to various semiconductor heterostructures with a similar design.
However, in the rest of the paper we will concentrate on a particular heterostructure for which a set of experimental data on time-resolved PL has been obtained allowing the comparison with the theoretical analysis \cite{KorenevSapega,Akimov2014}.
The width of the QW under consideration is $a=10$ nm and its depth is controlled by In composition. For $x_{\rm In}=10\%$ the QW depth for the electrons (i.e. the position of the first size quantization level relative to the GaAs conductance band edge) is $W_e=45$ meV for the temperature $T=2 K$.
The details of the structure design and fabrication can be found in Ref.\cite{Zaitsev}. The width $d$ of the spacer separating the Mn layer from the QW is varied in the range 2-10 nm for different samples and it is penetrable for the electrons as well as for the holes \cite{OURPRB2013}.
It is well known that Mn impurity in GaAs matrix can exist in two different states.
A single Mn atom replacing Ga atom in the lattice makes an Mn$_{Ga}$ configuration, where Mn behaves as
an acceptor with the hole binding energy $E_a\approx110$ meV.
Mn atom can also occupy an interstitial position Mn$_{I}$, at that it becomes a double-donor.
It has been confirmed experimentally that in the samples under study both Mn$_{Ga}$ and Mn$_{I}$ configurations
are realized \cite{Zaitsev}.
The acceptors Mn$_{Ga}$ provide weakly localized holes mediating the ferromagnetism. The holes are distributed between the Mn doping layer
and the QW. It was argued that both fractions can contribute to ferromagnetism depending on the QW depth \cite{ourAPL2015}
The samples being discussed in this paper had a very shallow QW for the holes ($W_h\approx30 \text{meV}$) so that the tunnel coupling between
the Mn layer and the QW is of non-resonant character. Thus, the equilibrium spin polarization of the holes located in the QW is not expected and (along with the kinetics discussed below) cannot fully explain the observed circular polarization of the photoluminescence from the QW.
The interstitial Mn$_I$ are known to be effective non-radiative recombination centers \cite{Nemec}.
Unlike Mn$_{Ga}$ substitutional impurity, the Mn$_{I}$ donors repel positively charged holes and do not directly participate in the hole-mediated ferromagnetism.

However, it was shown that there is a strong antiferromagnetic superexchange interaction between Mn ions in an interstitial and a neighboring substitutional
position \cite{Takeda2008,DietlReview}.
Thus, it is reasonable to assume that while there is a macrosopic magnetisation of the sample not only Mn$_{Ga}$ spins are ferromagnetically aligned but the
Mn$_{I}$ spins are aligned as well (in the opposite direction) \cite{Takeda2008}. With that taken into account we conclude that the electron bound states at Mn$_{I}$ ions
are split in spin projection on the same axis due to exchange interaction with the core $d^5$ electrons.
Now let us we consider an electron tunneling from the QW into the spin-split bound state at Mn$_I$ with the subsequent non-radiative recombination with
a valence band hole.
Taking into account the donor level spin splitting we note that there are two tunneling channels corresponding to the opposite electron spin projections.
The difference in the tunneling rates for spin-up and spin-down electrons tunneling from the QW to the donor states gives rise to a spin polarizations of the electrons remaining in the QW.
We argue that this mechanism is responsible for the observed polarization of the PL emitted from the QW.
Let us now proceed to a more detailed theoretical analysis of the phenomena.
The exchange interaction between the localized electron and Mn$_{I}$ core can be expressed as:
\begin{equation}
\label{eqVex}
V_{ex} = - \alpha_e  {\bf J} {\bf S},
\end{equation}
where ${\bf J}$ is the Mn$_I$ spin operator, ${\bf S}$ is electron spin operator, $\alpha_e$ is the exchange coupling constant.
The quantity $\alpha_e$, which describes the $s-d$ exchange is positive favouring the ferromagnetic alignment of the donor state electron with the $d^5$ core electrons.
Let $E_d$ be the donor energy level (measured from the size quantization level in the QW as shown in Fig.1).
Due to the exchange interaction (\ref{eqVex}) the level is split into two spin sublevels having the enerigies
\[
\varepsilon_{s} = E_d - s \Delta  ,\,\,\,\,\,\      \Delta =  \alpha_e J_z,
\]
where $s$ denotes the electron spin projection and $J_z$ is the Mn$_I$ spin projection onto $z$-axis directed normal to the QW plane.
In the external magnetic field applied along $z$ axis $B>0$ the ground state of substitutional Mn$_{Ga}$ corresponds to $-5/2$ spin projection, thus the Mn$_I$ center has the opposite spin  $J_z = + 5/2$. Consequently, the donor state  $s = +1/2$ has lower energy than $s=-1/2$.
The position of the donor level energy $E_d$ is quite a delicate question.
To the best of our knowledge there is no experimental data for the Mn$_I$ donor energy levels position. Theoretical calculations for bulk GaMnAs show the Mn$_I$ donor energy level to be lying in the conductance band \cite{Zunger}.
In this case because of the energy mismatch there is no direct resonant tunneling from the occupied electron states in the QW onto the spin split donor level. However, the spin splitting would manifest itself in a second order process involving the electron tunneling to the donor level followed by a phonon-assisted non-radiative recombination of the electron with the valence band hole in Mn layer.
The initial state corresponds to the electron 2D in the QW-plane with the energy $\varepsilon_k = \hbar^2 k^2/2 m$, where $m$ is the in-plane effective mass, and the final state is the electron in the valence bamd in the Mn layer and
%
an emitted phonon with the energy $\hbar \Omega_q$.
We neglect Zeeman splitting for 2D electrons in the QW in comparison with the exchange splitting at the donor site \cite{Zaitsev}.
The matrix element for the electron transition from the QW into the Mn layer is calcutated using the second order perturbation theory.
Using the Fermi's Golden Rule we get the transition rate:
\begin{equation}\label{eqTunRateSOQM}
d\omega_{{\bf k}, s}^{\bf q} = \frac{2\pi}{\hbar} \left| \frac{V_{q} T}{\varepsilon_k - E_d + s \Delta} \right|^2 \delta(\varepsilon_k+E_{rec}-\hbar \Omega_{\bf q})  \frac{S d^2{\bf k} d\nu_{q}}{(2\pi)^2},
\end{equation}
where $V_{q}$ is the electron-phonon interaction matrix element,
$d\nu_{q}$ is the phonon modes density, $E_{rec}$ is the energy difference between the electron size quantization level in the QW and the top of the valence band in the GaAs barrier (see Fig.\ref{fig1}), $T$ is the tunneling matrix element.
$T$ depends exponentially on spacer thickness $d$ as $T=\tau e^{- q d}$, where $q = \sqrt{ 2 m_{\ast} W_e /\hbar^2}$, $m_{\ast}$ is the perpendicular effective mass, $W_e$ is the electron barrier height (Fig.1), the pre-exponential factor is discussed in \cite{OURLowTempPhys}.
Integration of (\ref{eqTunRateSOQM}) over phonon degrees of freedom yields
\begin{equation}\label{eqTunRateSOQM_2}
d\omega_{{\bf k}, s} =   \frac{1}{\tau_{d}} \frac{\tau^2 e^{-2 q d}}{\left(\varepsilon_k - E_d + s \Delta \right)^2} \frac{S d^2{\bf k} }{(2\pi)^2},
\end{equation}
where we introduced the electron lifetime at the donor state
\[
\frac{1}{\tau_{d}} = \frac{2\pi}{\hbar} \int d\nu_{\bf q} |V_{q}|^2 \delta(E_{rec}-\hbar \Omega_{\bf q}).
\]
The dependence of $\tau_d$ on $k$ can be neglected as $\varepsilon_k\ll E_{rec}$.
Because the donor level is split the transition rate (\ref{eqTunRateSOQM})
appears to be different for the opposite spin projections.
The non-radiative current density for each spin projection is given by:
\begin{equation}\label{eqTunCur}
{j_s } = e\int {f_s \left( {{\bf{k}}} \right)\frac{d{\omega _{{\bf{k}},s}}}{S}},
\end{equation}
where the integration is performed over all 2D states and $f_s \left( {{\bf{k}}} \right)$ denotes the thermal distribution function of the electrons in the QW, $e$ is the electron charge.
The electron gas is non-degenerate since the electrons appear in the QW only due to photoexcitation. This implies:
\begin{equation}\label{eqFuncEl}
f_s({\bf k}) = \exp \left( {\frac{{\mu_s  - {\varepsilon _k}}}{{{k_B}{T_e}}}} \right),
\end{equation}
where $T_e$ is the electron temperature, $\mu_s$ is the chemical potential for the electrons with $s$ spin projection, $k_B$ is Boltzmann constant. The current density for each spin projection (\ref{eqTunCur}) is expressed as:
\begin{equation}\label{eqProm2}
j_{s} = e \frac{\tau^2 e^{-2qd}}{\tau_d} \frac{n_{s}}{k_B T_e} \int\limits_0^{\infty} \frac{ e^{-\varepsilon_k/k_B T_e }}{\left( \varepsilon_k - E_d +s \Delta \right)^2} d\varepsilon_k.
\end{equation}
As there is a large energy mismatch between the electrons in the QW and the donor level we have $k_B T_e \ll E_d$,
the expression (\ref{eqProm2}) can be simplified by setting $\varepsilon_k=0$ in the denominator,
then the final expression for the current reads:
\begin{equation}\label{eqTunFin_SO}
j_{s} = e \Gamma_{s} n_{\alpha} ,\,\,\,\,\,\,\,\ \Gamma_{s} =
 \frac{ e^{-2qd}}{\tau_d} \frac{\tau^2}{\left( E_d - s \Delta \right)^2},
\end{equation}
where $n_{s}$ is the sheet density of the electrons with $s$ spin projection in the QW, $\Gamma_{s}$ is the non-radiative channel recombination rate.
Note, that while $\Gamma_{s}$ (\ref{eqTunFin_SO}) is reduced by large value of $E_d$, the small carrier
lifetime at the donor level $\tau_d$ can keep the non-radiative current sufficiently high while the perturbation theory still holds.
Even with the equal sheet densities of spin-up and spin-down electrons, since $\Gamma_{+1/2} \neq \Gamma_{-1/2}$ there is a difference between spin-up and spin-down currents. For $\Delta \ll E_d$ this difference makes
\begin{gather}\label{eqSpinCur}
j_{+1/2} - j_{-1/2} = e n \frac{ e^{-2qd}}{\tau_d} \frac{ 2 \tau^2}{E_d^2} \frac{\Delta}{\varepsilon_d},
\end{gather}
where $n$ is the electron sheet density in the QW. In a positive external magnetic field $\Delta>0$, so $j_{+1/2}/e > j_{-1/2}/e$, so the $s=1/2$ channel is more efficient. This imbalance leads to accumulation of spin-down electrons in the QW.

\section{Comparison with experiment}
We applied the model described above to the experimental data on time-resolved PL \cite{KorenevSapega,Akimov2014}.
 As the transition rates for spin up and spin down electrons leaving the QW through the non-radiative channel
 are different ($\Gamma_{+1/2} \neq \Gamma_{-1/2}$), a nonzero spin accumulates in the QW.
The spin polarization of the electrons remaining in the QW gives rise to the circular polarization of the PL from the QW.
Let us define electron spin polarization degree as follows:
\begin{equation}
\rho_s = \frac{ n_{+1/2} - n_{-1/2}}{n_{-1/2}+n_{+1/2}}.
\end{equation}
$\rho_s$ is negative when $\alpha=-1/2$ electrons prevail. In this case
the radiative recombination with a heavy hole with the angular momentum projection $j_z=+3/2$ would be more efficient so the circular polarization $\sigma_+$ would dominate in the PL. The sign of the PL circular polarization (which is opposite to the sign of $\rho_s$ as defined above) is an important indicator confirming the applicability of the suggested spin polarization mechanism to explain the experimental observations. In our model the non-radiative transition rate is higher for spin-up electrons so that $\rho_s<0$ and $\sigma_+$ circular polarization dominates in the PL from the QW. This is exactly what is observed in the experiment.
Let us now focus on the kinetics of the photoexcited electrons.
The characteristic time of non-radiative recombination in (Ga,Mn)As is
$\tau_d \sim 1$ ps \cite{Nemec}.
However the electrons leave QW with the characteristic time $\Gamma^{-1}$ obtained from the  formula (\ref{eqTunFin_SO}), which is much longer than $\tau_d$. This fact indicates that the characteristic time of electron density decreasing in QW is governed by the tunneling process. The non-radiative
channel time $\Gamma^{-1}$
 in its turn, is much faster than the radiative recombination in the QW as confirmed by the calculations and experiment.
So we have the hierarchy of times $\tau_d \ll \Gamma^{-1} \ll \tau_{\rm rad}$. Another process which has a direct influence on $\rho_s$ is the carrier spin relaxation.
It was found that in the structures under study the holes spin relaxation time is very short $\tau_s^h \sim 10$ ps, while for the electrons it is much longer $\tau_s^e\sim 10$ ns and can be comparable to $\tau_{\rm rad}$ \cite{KorenevSapega}.

Let us now consider the PL dynamics in the time-resolved experiment. A short non-polarized excitation pulse ($\tau_{pulse} \approx 1$ ps \cite{KorenevSapega}) produces non-equilibrium carriers in the QW.
%
Because  $\Gamma^{-1} \ll \tau_{\rm rad}, \tau_s^e$,  right after the excitation pulse is switched off the PL characteristics are primarily governed by the tunneling.
Assuming immediate energy relaxation the electron dynamics is described by the following simple expression:
\[
\frac{d n_{s}}{dt} = - j_{s}/e.
\]
With $j_{s}$ given by (\ref{eqTunFin_SO}) we get that $n_{s}$ decays exponentially:
\[
n_{s} = \frac{n_0}{2} e^{- \Gamma_{\sigma} t },
\]
where $n_0$ is the electron sheet density generated in the QW by the excitation pulse.
From the known pulse duration the upper bound estimate is $n_0\sim 10^9$ cm$^{-2}$.
Then the density of the electrons in the QW and the spin polarization degree $\rho_s$ obey:
\begin{gather}
n(t) =
n_{+1/2} + n_{-1/2} =
\frac{n_0}2 (e^{-\Gamma_{+1/2} t} + e^{-\Gamma_{-1/2} t})\nonumber \\
\rho_s(t) = 
\tanh{\left(\frac{\delta \Gamma}2 t\right)}, \,\,\,\,\,
\delta \Gamma = \Gamma_{+1/2} - \Gamma_{-1/2}
\label{eqPL}
\end{gather}
\begin{figure}[h]
 \centering\includegraphics[width=0.45\textwidth]{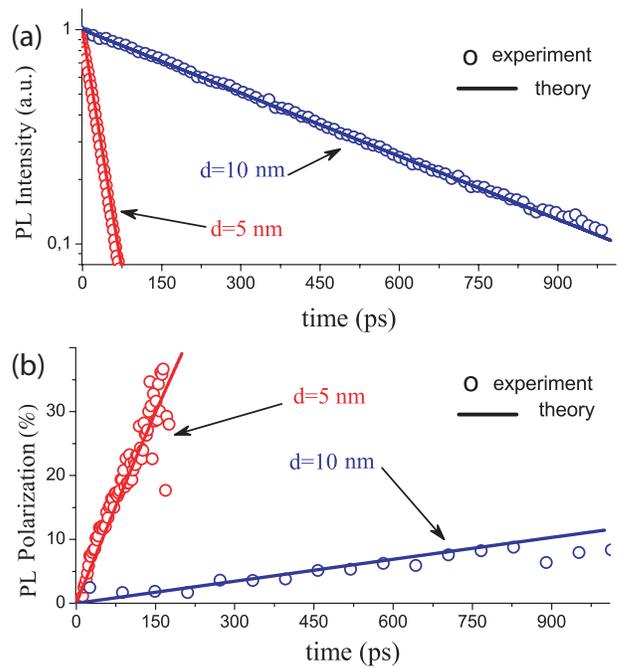}
 \caption{Time dependence of the intensity (a) and the circular polarization degree (b) of the photoluminescence
 from the QW}
\label{fig3}
\end{figure}

Shown in Fig. 2 is a comparison of the calculated electron sheet density and spin polarization dynamics (\ref{eqPL}) with the time dependence of the PL observed in experiments \cite{KorenevSapega,Akimov2014} for two different spacer thickness $d=5,10$.
In the calculation we used the following parameters: $\tau= 10$ meV, $\tau_d = 0.1$ ps, $E_d = 47$ meV, $W_e= 45$ meV, $\Delta = 2.5$ meV.
Fig.2 (a) shows the PL intensity decreasing with time following the kinetics of the photoexcited electrons.
As noted above the decay of the PL intensity is very fast compared to the radiative recombination time $\tau_{\text{rad}}$.
The strong dependence of the PL on the spacer thickness is due to the exponential tunneling factor in (\ref{eqTunFin_SO}).
The PL circular polarization degree is shown in Fig.2 (b).
The spin-dependent tunneling ($\delta \Gamma\neq 0$) and long electron spin relaxation time $\tau_s^e$ leads to the accumulation of nonequilibrium spin in the QW $\rho_s$, which increases linearly with time while $t\ll\delta\Gamma^{-1}$ (\ref{eqPL}).
It should be noted that the electron tunneling also leads to accumulation of the positive charge in the QW which might have prevented
the tunneling due to electrostatic effect.
However, the value of the positive charge is of the order of the initial electron concentration right after photoexcitation $n_0$,
i.e. no more than  $n_0 \sim 10^9$ cm$^{-2}$, such a charge density is too small to significantly affect the tunneling by electrostatic effects.
\section{Summary}
To conclude, we proposed a microscopic mechanism explaining the observed
ultrafast PL kinetics and the PL circular polarization in
hybrid ferromagnetic-semiconductor structure with a QW and spatially separated Mn $\delta$-layer.
The PL behavior in our model is governed by the dynamics of the photoexcited electrons.
 The key process which determines the electrons dynamics is their tunneling onto Mn$_I$ interstitial donor-like defects in the Mn layer.
The spin splitting of this donor in the exchange field of the Mn layer makes this tunneling spin-dependent and causes accumulation of nonequilibrium spin in the QW. While the tunneling itself is of a non-resonant character, the spin-dependent carriers leakage through the non-radiative channel
appears to be effective due to the very small lifetime of a carrier at the donor site.
 This proposed model allowed us to explain the fast PL decay,
 linear increase of the PL circular polarization with time as well as its sign observed in the experiment.
\section{Acknowledgments}
We thank I. A. Akimov for kindly providing experimental data, V. L. Korenev, V. F. Sapega and S. V. Zaitsev for very fruitful discussions.
This work was supported by the Government of Russia through the program P220 (project
14.Z50.31.0021, leading scientist M. Bayer). K.S.D. acknowledges the support of Dynasty foundation.
\bibliography{FanoExchange}

\end{document}